\documentclass[aps,prb,twocolumn,floats,showpacs]{revtex4}
\usepackage{epsfig}
\usepackage{amsmath}
\usepackage[dvips]{color}
\usepackage{amssymb}
\usepackage{amsopn}
\usepackage{amstext}
\usepackage{amsbsy}
\usepackage{amscd}
\usepackage{amsxtra}

\usepackage{bm}

\def\up{\uparrow}
\def\down{\downarrow}

\newcommand{\veps}{\varepsilon}
\newcommand{\bfk}{{\bf k}}

\newcommand{\ua}{\uparrow}
\newcommand{\da}{\downarrow}
\begin{document}

\title{Anderson Impurity in the Bulk of Topological Insulators}

\author{Igor Kuzmenko$^1$, Yshai Avishai$^{1,2}$ and Tai Kai Ng$^2$}
\affiliation{\footnotesize
  $^1$Department of Physics, Ben-Gurion University of
  the Negev Beer-Sheva, Israel \\
  $^2$Department of Physics, Hong Kong University of Science and
  Technology, Kowloon, Hong Kong}

\begin{abstract}
When an Anderson impurity is immersed in the bulk of a topological
insulator that has an "inverted-Mexican-hat"  band-dispersion, a
Kondo resonant peak appears simultaneously with an in-gap
bound-state. The latter generates another spin state thereby
screening the Kondo effect. Using weak-coupling RG scheme,  it is
shown system exhibits complex crossover behavior  between
different symmetry configurations and may evolve into a
self-screened Kondo or an $SO(n), (n=3,4)$ low energy fixed
points. Experimental consequences of these scenarios are pointed
out.
\end{abstract}


\maketitle

\section {Introduction}
  \label{sec-intro}

The significance of topological insulators (TI) as a new state of
matter has been stressed in  numerous publications.%
\cite{1,2,3,4,5,6,7} So far, the main attention has been focused
on the surface states.\cite{8,9,10,11,12,13,14,15,16,17,18,19,20,%
21,22,22a, 22b} It was pointed out recently that impurity
scattering may have non-trivial effects in the {\it bulk} of TI
due to its peculiar band structure.\cite{23,24,top-ins-3D-10,%
3D-TI-QD-12} More concretely, in topological insulators with
''inverted-Mexican-hat" band dispersion around the $\Gamma$-point,
in-gap bound states occur due to impurities from which band
electrons are scattered (even with arbitrary weak scattering
rate).\cite{23} In the case of magnetic (Anderson) impurities
(forming localized moments) the associated Kondo physics in bulk
TI is profoundly distinct from its metallic analog. Using a
slave-boson mean-field theory,\cite{24} it was shown that an
antiferromagnetic exchange interaction $\sim{J}_{df}{\bf{S}}_d%
\cdot{\bf{S}}_f$ (with $J_{df}>0$) between the spins of the
Anderson impurity $d$ and the induced mid-gap bound-state $f$,
leads to a self-screened Kondo effect (KE).\cite{24} In fact, the
physics described above is not limited to TI but is a general
consequence of insulators (and semi-conductors) with a large
electronic density of states at the band edge such that in-gap
bound states are easily induced by an Anderson impurity.

The goal of this work is to perform a detailed analysis of the
interplay between the Anderson impurity and its induced in-gap
bound state in bulk TI, and, in particular, elucidate the
observable quantities (such as conductivity and magnetic
susceptibility) related to the underlying Kondo physics. Employing
a weak-coupling renormalization group (RG) analysis, we show that
the system exhibits complex crossover behavior between different
symmetry sectors and that the exchange interaction $J_{df}$
between the $d$- and $f$-spins may be renormalized dynamically to
a negative value at suitable parameters regime. In this case the
KE is not screened and the low energy physics of the system is
described by an SU(2), SO(4) or SO(3) Kondo fix point. The
temperature dependence of the impurity induced resistance and
magnetic susceptibility are studied at various regimes.

The article is organized as follows. In Section \ref{sec-model},
we introduce the model considered starting from a bare Anderson
type Hamiltonian. It is shown  that the presence of an Anderson
impurity $d$ leads to potential scattering of the conduction band
electrons,  that, in turn, brings about   the formation of an
in-gap localized quantum state $f$. We introduce the singlet and
triplet states of the composite $(d+f)$ impurity. At the end of
Section \ref{sec-model} we arrive at an effective Anderson
Hamiltonian that will be analyzed further in the next sections.
The local density of states (DOS) is calculated in Section
\ref{sec-local-DOS} for the TI with the potential scattering term.
Renormalization group analysis is carried out in Section
\ref{sec-RG}. Depending on various energy domains, the RG analysis
ends up with various Kondo Hamiltonians that posses different
(dynamical) symmetries, SU(2), SO(3) or SO(4). Section
\ref{sec-resist-suscept} is devoted to the calculations of
electric resistivity and magnetic susceptibility in the relevant
energy domains. One of the central results of our work is the
occurrence of temperature driven crossovers of KE between
different symmetry classes. The results are then summarized in
Section \ref{sec-conclusions}. Some details of the calculation
technique are presented in the appendices. These include physical
origin of the potential scattering induced by an Anderson impurity
(appendix \ref{append-what-is-V0}), discussion of local density of
states (appendix \ref{append-local-DOS}), the SO(4) Kondo
Hamiltonian (appendix \ref{append-HK-SO4}), and finally,
calculations of electric resistivity (Appendix
\ref{append-resist}) and magnetic susceptibility (Appendix
\ref{append-suscept}) for different temperature intervals.

\section{\bf Model Hamiltonian}
  \label{sec-model}
Our aim in this section is to derive an effective Hamiltonian
that, in addition to the Anderson  impurity $d$, contains also the
mid-gap state $f$ as discussed in Ref [\onlinecite{24}]. Starting
from an Hamiltonian describing an Anderson impurity $d$ in the
bulk of a TI, a few manipulations are required to transform it
into its workable form, Eq.~(\ref{H-model-def}) below. The bare
Hamiltonian is,
\begin{eqnarray}
  &&
  H=H_0+H_d+H_t^{(0)}+V.
  \label{H-tot-def}
\end{eqnarray}
Here the first term, $H_0$, describes electrons in the bulk of the
TI,
\begin{eqnarray}
  &&
  H_0=
  \sum_{\bf{k}}
  \Psi^\dagger_{\bf{k}}~
  h_0({\bf{k}})~
  \Psi_{\bf{k}},
  \nonumber
  \\
  &&
  h_0(\bfk) =
  \hbar v_F
  \big(
      {\bf{k}}
      \cdot
      \boldsymbol\alpha
  \big)+
  \beta
  M_k,
  \label{H-ti-def}
  \\
  &&
  \Psi^\dagger_\bfk=
  \Big(
      a^{\dagger}_{\bfk \ua},
      a^\dagger_{\bfk \da},
      b^\dagger_{\bfk \ua},
      b^\dagger_{\bfk \da}
  \Big).
  \nonumber
\end{eqnarray}
where $M_k=mv_F^2-B \hbar^2 k^2$, $a^\dagger_{\bfk \sigma}$,
$b^\dagger_{\bfk \sigma}$ are creation operators for electron of
momentum $\bfk$ and spin projection $\sigma$. $h_0(\bfk)$ is
written in particle-hole $\otimes$ spin space, with
$\boldsymbol\alpha=t_x\otimes{\bf{s}}$ and
$\beta=t_z\otimes{s_0}$, where ${\bf{t}}=(t_x,t_y,t_z)$ or
${\bf{s}}=(s_x,s_y,s_z)$ are vectors of the Pauli matrices acting
in the space of isospins or spins, $s_0$ is the $2\times2$
identity matrix. $H_d$ is the Hamiltonian for the Anderson
impurity,
\begin{equation}
  \label{hd}
  H_d=
  \epsilon_d
  \sum_{\sigma}
  n_{d\sigma}+
  U_dn_{d\up}n_{d\down},
\end{equation}
where $\epsilon_d$ is the impurity energy level and $U_d$ is the
interaction between electrons on the impurity.
$n_{d\sigma}=d_{\sigma}^{\dag}d_{\sigma}$, $d_{\sigma}^{\dag}$ or
$d_{\sigma}$ is the creation or annihilation operator of electron
on the d-level. The hybridization between the Anderson impurity
$d$ and the band electrons is described by
\begin{equation}
 \label{ht}
  H_t^{(0)} =
  V_d
  \sum_{{\bf{k}},\sigma}
  \left(
      a_{{\bf{k}}\sigma}^{\dag}
      d_{\sigma}-
      i b_{{\bf{k}}\sigma}^{\dag}
      d_{\sigma}+
      {\text{H.c.}}
  \right).
\end{equation}

The last term appearing in the Hamiltonian (\ref{H-tot-def}) is an
effective potential scattering between band-electrons induced by
the impurity,
\begin{equation}
  V=
  V_0
  \sum_{{\bf{k}},\bf{k}'\sigma}
  \Big(
      a_{{\bf{k}}\sigma}^{\dag}-
      i b_{{\bf{k}}\sigma}^{\dag}
  \Big)
  \Big(
         a_{{\bf{k}}'\sigma}+
         i b_{{\bf{k}}'\sigma}
  \Big).
  \label{V-def}
\end{equation}
A few words on the origin of the potential scattering term $V$ are
in order. When an Anderson impurity is immersed in a metal it
leads to an $s-d$ Hamiltonian and to a potential scattering. In
the standard analysis of the Anderson impurity, the induced
potential scattering term is neglected because it is irrelevant as
far as Kondo physics is concerned. The situation is different when
the impurity is immersed in an insulator. Here, as we shall see in
the following, the potential scattering leads to a mid gap
bound-states that profoundly affects the underlying physics.  The
detailed discussions about the physical origin of the strength
$V_0$ and the derivation of the localized in-gap $f$-level are
given in Appendix \ref{append-what-is-V0}.

\noindent
{\bf Diagonalization of $H_0$}\\
The eigenstates of $H_0$ alone is given by
\begin{subequations}
  \label{bulk-disp}
\begin{eqnarray}
  H_0 &=&
  \sum_{\nu\sigma{\bf{k}}}
  \nu\veps_{k}
  \gamma_{\nu{\bf{k}}\sigma}^{\dag}
  \gamma_{\nu{\bf{k}}\sigma},
  \label{H-ti-diag-append}
\end{eqnarray}
where $\nu=\pm 1$ or $c/v$ denotes the conduction and valence
band.
\begin{equation}
  \label{disp}
  \veps_{k}=
  \sqrt{M_{k}^2+(\hbar v_F k)^2},
  \ \ \
  M_{k}=m v_F^2-B \hbar^2k^2,
\end{equation}
is the band dispersion. $\gamma_{\nu{\bf{k}}\sigma}^{\dag}$ and
$\gamma_{\nu{\bf{k}}\sigma}$ are creation and annihilation
operators of quasi-particle defined through the transformation,
\begin{eqnarray}
  &&
  \tilde\Psi_{\bf{k}} =
  {\cal{S}}_{k}
  {\cal{U}}_{k}
  \Psi_{\bf{k}},
  \ \ \ \ \
  \tilde\Psi_{\bf{k}}^{\dag}=
  \Big(
      \gamma_{c{\bf{k}}\ua},
      \gamma_{c{\bf{k}}\da},
      \gamma_{v{\bf{k}}\ua},
      \gamma_{v{\bf{k}}\da}
  \Big),
  \nonumber
  \\
  &&
  {\cal{U}}_{k} =
  \cos\frac{\alpha_k}{2}+
  i \sin\frac{\alpha_k}{2}~
  t_y \otimes ({\bf{s}}\cdot{\bf{e}}_{k}),
  \label{transform-def}
\end{eqnarray}
where ${\bf{e}}_k={\bf{k}}/k$,
$$
  \tan\Big(\frac{\alpha_{k}}{2}\Big) =
  \sqrt{\frac{\veps_{k}-M_{k}}
             {\veps_{k}+M_{k}}}.
$$
${\cal{S}}_{k}$ is a unitary matrix which commute with
$h_0({\bf{k}})$ and ${\cal{U}}_{k}$.
\end{subequations}
$\veps_k$ is gapped and the insulator is topological for $Bm>0$.
For $Bm>1/2$ (assumed hereafter), the band dispersion has an
''inverted-Mexican-hat" form (see Figure 1) with dispersion
minimum at a surface of nonzero wave-vector $\mathbf{q}$'s, with
\begin{eqnarray}
  \veps_q =
  \frac{v_F^2}{2B}~
  \sqrt{4 B m-1},
  \ \ \ \ \
  q=
  \frac{v_F}{\hbar B}~
  \sqrt{B m-\frac{1}{2}},
  \label{q-veps-q-def}
\end{eqnarray}
where $q=|\mathbf{q}|$.

\begin{figure}[htb]
\centering
\includegraphics[width=60mm,angle=0]{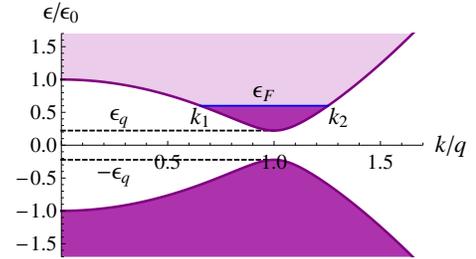}
\caption{\footnotesize
  Energy dispersion (\ref{disp}) for $Bm=20$. The dark
   (bright) regions denote the energy levels below (above)
   the Fermi energy $\epsilon_F$. The band-edges occur
   at momentum $q$  at energies $\pm \veps_q$.
   (Here it is assumed that
   $\veps_q<\epsilon_F<mv_F^2$). $k_1$ and $k_2$ are two
   solutions of the equation $\veps_k=\epsilon_F$.}
 \label{Fig1}
\end{figure}

\noindent
{\bf Transformation of $H_t^{(0)}$ and $V$}\\
Here we express the terms $H_t^{(0)}$ and $V$ defined in equations
(\ref{ht}) and (\ref{V-def}) in terms of the quasiparticle
operators. Applying the unitary transformation
(\ref{transform-def}) to the tunnel Hamiltonian (\ref{ht}), we get
\begin{eqnarray*}
  H_t &=&
  \sum_{\bf{k}}
  \Bigg\{
       \tilde\Psi_{\bf{k}}^{\dag}
       {\cal{S}}_{k}
       \bigg(
       \begin{array}{c}
       {\cal{V}}_{k} \\ -i{\cal{V}}_{k}
       \end{array}
       \bigg)
       \bigg(
       \begin{array}{c}
       d_{\ua} \\ d_{\da}
       \end{array}
       \bigg)+
       {\rm{H.c.}}
  \Bigg\},
\end{eqnarray*}
where ${\cal{V}}_{k}$ is a $2\times2$ unitary matrix,
\begin{eqnarray*}
  {\cal{V}}_{k} &=&
  \cos\alpha_k-
  i\sin\alpha_k
  ({\bf{s}}\cdot{\bf{e}}_{k}),
  \ \ \ \ \
  {\cal{V}}_{k}^{\dag}
  {\cal{V}}_{k} = 1.
\end{eqnarray*}
Choosing ${\cal{S}}_{k}$ as,
\begin{eqnarray}
  {\cal{S}}_{k} &=&
  \left(
  \begin{array}{cc}
  {\cal{V}}_{k}^{\dag} & 0 \\
  0 & i{\cal{V}}_{k}^{\dag}
  \end{array}
  \right),
  \label{S-matr-def}
\end{eqnarray}
we get,
\begin{equation}
  \label{hteff}
  H_t = V_d\sum_{{\bf{k}},\nu,\sigma}
  \left(\gamma_{\nu {\bf{k}}\sigma}^{\dag}
      d_{\sigma}+d^{\dag}_{\sigma}\gamma_{\nu {\bf{k}}\sigma}
  \right).
\end{equation}

Applying transformations (\ref{transform-def}) and
(\ref{S-matr-def}) to the potential scattering Hamiltonian
(\ref{V-def}), we get,
\begin{eqnarray*}
  V &=&
  V_0
  \sum_{\nu\nu'{\bf{kk}}'\sigma}
  \gamma_{\nu{\bf{k}}\sigma}^{\dag}
  \gamma_{\nu'{\bf{k}}'\sigma}.
\end{eqnarray*}

\noindent
{\bf Diagonalization of $H_0+V$: Mid-gap state.}\\
The next step is crucial, as it demonstrates the formation of a
mid-gap state $f$ due to the potential scattering term $V$.
Because the potential $V$ emerges due to the Anderson impurity, we
say that the mid-gap state is induced by the impurity. In order to
find the eigenstates of $H_0+V$, we solve the Heisenberg equation
of motion
$$
  i\hbar\dot\gamma_{\nu{\bf{k}}\sigma} =
  \big[H_0+V,\gamma_{\nu{\bf{k}}\sigma}\big].
$$
Taking $\gamma_{\nu{\bf{k}}\sigma}(t)=\gamma_{\nu{\bf{k}}\sigma}%
e^{-i\epsilon t/\hbar}$, we get
\begin{eqnarray}
  \epsilon
  \gamma_{\nu{\bf{k}}\sigma}
  &=&
  \nu\veps_{k}
  \gamma_{\nu{\bf{k}}\sigma}+
  V_0
  \sum_{\nu'{\bf{k}}'}
  \gamma_{\nu'{\bf{k}}'\sigma}.
  \label{eq-Heisenberg}
\end{eqnarray}
Equation (\ref{eq-Heisenberg}) has nontrivial solution when
$\epsilon$ satisfies the secular equation,
\begin{eqnarray}
  \sum_{\nu{\bf{k}}}
  \frac{V_0}{\epsilon-\nu\veps_{k}}
  &=& 1.
  \label{eq-secular}
\end{eqnarray}
The solutions for $|\epsilon|\ge\veps_q$ describe the band
electrons, whereas the solution for $|\epsilon|<\veps_q$
corresponds to the localized $f$-level.

It is seen that when $|\epsilon|<\veps_q$, the expression in the
left hand side of the secular equation (\ref{eq-secular}) is
positive when $\epsilon$ is negative and negative when $\epsilon$
is positive [$V_0$ is assumed to be positive here]. When
$|\epsilon|\to\veps_q$, the sum diverges as
$\big(\veps_q^2-\epsilon^2\big)^{-1/2}~{\rm{sign}}(-\epsilon)$  for
energy bands with the ``inverted Mexican hat" structure.\cite{24}
As a result, the secular equation gives us a mid-gap energy level
$\epsilon_f$ which lies within the interval
$-\veps_q<\epsilon_f<0$. When $\epsilon_f+\veps_q\ll\veps_q$, we
can write
\begin{equation}
  \epsilon_f \approx
  -\veps_q~
  \sqrt{1-\Big(\frac{V_0 q}{2 \pi \hbar^2 B}\Big)^2}.
  \label{epsilon-f}
\end{equation}

This procedure leads to a minor modification
of the annihilation and creation operators for the band
electrons. Strictly speaking they are respectively expressed as
linear combination of $\gamma_{\nu{\bf{k}}\sigma}$ and
$\gamma_{\nu{\bf{k}}\sigma}^{\dag}$ using perturbation theory
with $V_0$ as a small parameter. However, since the potential pulls only one level (out of many)
from the band into  the gap,
the other levels are virtually unaffected. In what follows, we assume that the operators
for the modified levels inside the band
just slightly differs from the $\gamma_{\nu{\bf{k}}\sigma}$ and
$\gamma_{\nu{\bf{k}}\sigma}^{\dag}$ .  Thus, the main outcome of the potential scattering
is the creation of a midgap level $f$. The annihilation
operator $f_{\sigma}$ for this localized level is given by,
\begin{eqnarray}
  &&
  f_{\sigma} =
  \sum_{\nu{\bf{k}}}
  \frac{AV_0}{\epsilon_f-\nu\veps_k}~
  \gamma_{\nu{\bf{k}}\sigma},
  \label{f-gamma-def}
  \\
  &&
  A =
  \Bigg(
       \sum_{\nu{\bf{k}}}
       \frac{V_0^2}
            {\big(
                 \epsilon_f-
                 \nu\veps_q
             \big)^2}
  \Bigg)^{-\frac{1}{2}}.
  \nonumber
\end{eqnarray}

\noindent
{\bf{Hybridization term $H_{df}$}}\\
The last ingredient in our quest for constructing an effective
tunneling Hamiltonian with $d$ and $f$ impurities is to identify a
hopping term $H_{df}$ between the Anderson impurity and the
mid-gap state.  The hybridization term $H_t^{(0)}$ between the
bulk TI electrons and the $d$-impurity level leads to an effective
tunneling term $H_{df}$ between the $d$ impurity and the $f$
in-gap bound state, and an effective tunneling term $H_t$ between
the $d$ impurity and the band states. We assume that $H_t$ is
still given by equation (\ref{hteff}), whereas $H_{df}$ is
\begin{eqnarray}
  H_{df} &=&
  V_{df}
  \sum_{\sigma}
  \left(
       f^{\dag}_{\sigma}
       d_{\sigma}+
       d^{\dag}_{\sigma}
       f_{\sigma}
  \right),
  \label{Hdf-def}
\end{eqnarray}
where ${V}_{df}\sim{V}_{d}$.

\noindent
{\bf The effective Hamiltonian}\\
Finally, to arrive at the desired effective tunneling Hamiltonian
we collect all pieces into a sum of  three parts structured as
``band+composite impurity+hybridization" Hamiltonians,
\begin{subequations}
\begin{eqnarray}
  H &=&
  H_l+H_C+H_t.
  \label{H-model-def}
\end{eqnarray}
Here $H_l=H_0+V$ is the Hamiltonian of the two band electrons that
include also the potential scattering term, and
\begin{eqnarray}
  H_C &=&
  H_d+H_f+H_{df},
  \label{HC-def}
\end{eqnarray}
is the Hamiltonian of the composite impurity, including the $d$-
and $f$-levels. $H_d$ is given by equation (\ref{hd}), $H_{df}$ is
given by equation (\ref{Hdf-def}), and
\begin{eqnarray}
  H_f &=&
  \epsilon_f
  \sum_{\sigma}
  n_{f\sigma}+
  U_f
  n_{f\up}
  n_{f\down},
  \label{Hf-def}
\end{eqnarray}
where $\epsilon_f$ is the single-electron energy (\ref{epsilon-f})
of the localized $f$-level, $U_f$ is the interaction between
electrons on the $f$-level, and
$n_{f\sigma}=f_{\sigma}^{\dag}f_{\sigma}$. The hybridization term
$H_t$ is given by equation (\ref{hteff}).
  \label{subeqs-H-model}
\end{subequations}

\noindent
{\bf Energy scales}\\
Few words about energy scales are in order: Unless otherwise
specified, we shall assume $U_d\rightarrow\infty$ and
\begin{equation}
  \label{escale}
  \epsilon_F-D_0 <
  \epsilon_d \ll
  \epsilon_f <
  \epsilon_F <
  \epsilon_f+U_f\ll\epsilon_F+D_0,
\end{equation}
where $D_0$ (the initial bandwidth) is the highest energy cutoff,
and $\epsilon_F$ is the Fermi energy (see Figure \ref{Fig1}). We
use $\epsilon_F=2\veps_q$, $\epsilon_f\approx-\veps_q$,
$U_f=5\veps_q$ and $\epsilon_d=-80\veps_q$ in the following
calculations.

\noindent
{\bf The eigenstates of $H_C $}\\
 Equation  (\ref{HC-def}) are
specified by the configuration numbers $(N_d,N_f)$ representing
the number of electrons on the levels $d$ and $f$. With energy
scales specified by\ (\ref{escale}) the ground state has
$N_d=N_f=1$ and there are four possible states, a spin-singlet
state $|S\rangle$ and three spin-triplet states $|T_m\rangle$
($m=0,\pm1)$. The singlet energy is modified when $V_{df}\ne0$
while the triplet energy is unaffected. Explicitly,
\begin{subequations}
\begin{eqnarray}
  &&
  |S\rangle =
  \left\{
       \frac{\alpha_S}{\sqrt{2}}
       \Big(
           d_{\up}^{\dag}
           f_{\down}^{\dag}-
           d_{\down}^{\dag}
           f_{\up}^{\dag}
       \Big)-
       \beta_f
       f_{\up}^{\dag}
       f_{\down}^{\dag}
  \right\}
  |0\rangle,
  \label{singlet-def}
  \\
  \nonumber
  \\
  &&
  \begin{array}{l}
  \displaystyle
  |T_1\rangle=
  d_{\up}^{\dag}
  f_{\up}^{\dag}
  |0\rangle,
  \ \ \ \ \
  |T_{-1}\rangle=
  d_{\down}^{\dag}
  f_{\down}^{\dag}
  |0\rangle,
  \\
  \\
  \displaystyle
  |T_{0}\rangle =
  \frac{1}{\sqrt{2}}
  \Big\{
      d_{\up}^{\dag}
      f_{\down}^{\dag}+
      d_{\down}^{\dag}
      f_{\up}^{\dag}
  \Big\}
  |0\rangle,
  \end{array}
  \label{triplet-def}
\end{eqnarray}
  \label{subeqs-singlet-triplet}
\end{subequations}
where
\begin{eqnarray*}
  &&
  \varepsilon_{S} =
  \epsilon_d+
  \epsilon_f-
  \frac{2V_{df}^2}{\Delta_f},
  \ \ \ \ \
  \varepsilon_{T} =
  \epsilon_d +
  \epsilon_f,
  \\
  &&
  \alpha_S =
  \sqrt{1-\beta_f^2},
  \ \ \ \ \
  \beta_f =
  {\sqrt{2}V_{df}\over\Delta_f},
  \\
  &&
  \Delta_f =
  \epsilon_f-
  \epsilon_d+
  U_f.
\end{eqnarray*}
In the absence of hybridization of d-electron with the band
electrons (i.e., when $V_d=0$), $\veps_S<\veps_T$, and the spin
singlet state has lower energy.

\section{Local density of states}
  \label{sec-local-DOS}

Because of the potential scattering, the local density of states
(DOS) depends on the energy $\veps$ of an electron and on the
distance $r$ from the impurity,
\begin{eqnarray}
  &&
  \rho(\veps,r)
  =
  -\frac{1}{\pi}~
  {\rm{Im}}
  \sum_{\nu\nu'{\bf{k}}{\bf{k'}}}
  G_{\nu{\bf{k}},\nu'{\bf{k}}'}(\veps)
  e^{i({\bf{k}}-{\bf{k}}'){\bf{r}}},
  \label{DOS-local-def}
  \\
  &&
  G_{\nu{\bf{k}},\nu'{\bf{k}}'}(\veps)
  =
  \int
  \tilde{G}_{\nu{\bf{k}},\nu'{\bf{k}}'}(t)
  e^{\frac{(i\veps-\eta)t}{\hbar}}
  dt.
  \label{G-nuk-veps-def}
\end{eqnarray}
Here $\eta$ is a positive infinitesimal parameter,
$\tilde{G}_{\nu{\bf{k}},\nu'{\bf{k}}'}(t)$ is a retarded Green's
function,
\begin{eqnarray}
  \tilde{G}_{\nu{\bf{k}},\nu'{\bf{k}}'}(t) =
  -\frac{i}{\hbar}~
  \Theta(t)~
  \Big\langle
      \Big\{
          \gamma_{\nu{\bf{k}}\sigma}(t),
          \gamma_{\nu'{\bf{k}}'\sigma}^{\dag}
      \Big\}
  \Big\rangle,
  \label{G-nuk-t-def}
\end{eqnarray}
where $\langle\cdots\rangle$ denotes the thermal average with
respect to the Hamiltonian $H_0+V$.

Applying the Heisenberg equation of motion (\ref{eq-Heisenberg}),
we get the following expression for the local DOS,
\begin{eqnarray}
  \rho(\veps,r) &=&
  \rho_0(\veps)~
  \frac{1+
        \pi^2
        V_0^2~
        \big[
            \rho_0^2(\veps)-
            \rho_0^2(\veps,r)
        \big]}
            {1+
             \pi^2
             V_0^2~
             \rho_0^2(\veps)},
  \label{DOS-local-res}
\end{eqnarray}
where
\begin{widetext}
\begin{eqnarray}
  &&
  \rho_0(\veps,r) =
  \Theta(|\varepsilon|-\varepsilon_q)~
  \frac{\rho_c~|\varepsilon|}
       {2\sqrt{\varepsilon^2-\varepsilon_q^2}}~
  \Bigg\{
      \Theta(\veps_0-|\veps|)~
      \frac{\sin\big(qrg_1(\veps)\big)}{qr}+
      \frac{\sin\big(qrg_2(\veps)\big)}{qr}
  \Bigg\},
  \ \ \ \ \
  \rho_0(\veps) =
  \rho_0(\veps,0).
  \label{DOS-bulk-res}
\end{eqnarray}
\end{widetext}
 Detailed derivation of the local DOS is given in Appendix
\ref{append-local-DOS}.

\begin{figure}
\centering
\includegraphics[width=60mm, angle=0]
  {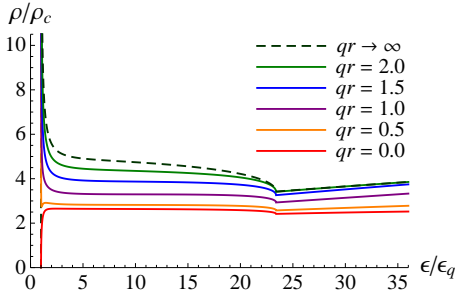}
 \caption{\footnotesize
   Local DOS (\ref{DOS-local-res}) for $\veps_0=24\veps_q$,
   $V_0\rho_c=0.06$ and different values of $qr$.}
 \label{Fig-DOS-local}
\end{figure}

The local DOS (\ref{DOS-local-res}) is shown in Figure
\ref{Fig-DOS-local} for different values of $qr$. It is seen that
$\rho(\epsilon,0)$ vanishes as
$\big(\epsilon^2-\epsilon_q^2\big)^{\frac{1}{2}}$ when
$|\epsilon|\to\epsilon_q$. When $qr>2$, $\rho(\epsilon,r)$
approaches $\rho_0(\epsilon)$, the bare DOS.

\section{Renormalization-Group (RG) Analysis}
  \label{sec-RG}
Within RG analysis, high energy charge fluctuations are integrated
out and one reaches a low energy spin ($s-d$)  Hamiltonian. The
latter is written in terms of the spin operator ${\bf s}$ of the
band electrons and a collection of vector operators of the
composite impurity. These vector operators generate a dynamical
symmetry group that characterize the pertinent Kondo
physics.\cite{28} If the ground state of $H_C$ contains a single
electron the symmetry group is SU(2), while if the ground state of
$H_C$ contains two electrons the symmetry group can be SO(3) [see
equation (\ref{HK-triplet})] or SO(4) [see equation
(\ref{HK-singlet-triplet})]. Technically,  the high energy degrees
of freedom are successively integrated out such that
$D_0\rightarrow D<D_0$ is reduced and energies are renormalized.
For the Kondo effect in metals, where the DOS is virtually
constant, this is a standard procedure. The fact that
$\rho(\veps)\ne$const. requires some modifications. We introduce
the following definitions,
\begin{eqnarray}
  \label{rhoav}
  &&
  \bar {\rho}(D) =
  \frac{1}{2}~
  \Big[
      \rho(\epsilon_F+D)+
      \rho(\epsilon_F-D)
  \Big],
  \label{L-def}
  \\
  &&
  {\cal L}(D_1,D_2)=
  \int\limits_{D_1}^{D_2}
  \frac{2~\bar{\rho}(D)~dD}{\rho_c~D}.
\end{eqnarray}

\begin{figure}[htb]
\centering
\includegraphics[width=60mm,angle=0]
  {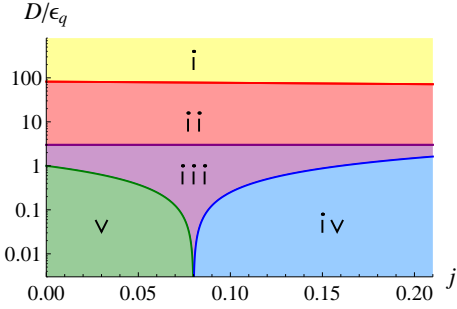}
 \caption{\footnotesize
   Parametric diagram $j-D$ for
   $\epsilon_d=-80\veps_q$, $\veps_0=24\varepsilon_q$ and
   $\epsilon_F=2\veps_q$. The curves separate the different
   temperature intervals:
   the red curve is $D_i=\epsilon_F-\veps_d$
   separating the regimes (\emph{i}) [mixed valence regime] and
   (\emph{ii}) [SU(2) Kondo regime],
   the purple line is $D_{ii}$ separating the regimes
   (\emph{ii}) [SU(2) Kondo regime] and \emph{iii} [SO(4) Kondo
   regime], the blue curve is $D_{iii}=|J_{df}|$ [$J_{df}<0$]
   separating the regimes (\emph{iii}) [SO(4) Kondo regime] and
   (\emph{iv}) [SO(3) Kondo regime], whereas
   the green curve is $D_{iii}=|J_{df}|$ [$J_{df}>0$] separating
   the regimes (\emph{iv}) [SO(4) Kondo regime] and
   (\emph{v}) [self screened Kondo regime].}
 \label{Fig-ParaDiagr}
\end{figure}

\noindent The RG flow is divided into the following regimes (see
Figure \ref{Fig-ParaDiagr}):
\begin{itemize}
\item[(\emph{i})]   $\epsilon_d>\epsilon_F-D>\epsilon_F-D_0$, in which
             charge fluctuations in both $d$ and $f$ states exist
             and there is no KE;
\item[(\emph{ii})]  $\epsilon_F-D>\epsilon_d$ but
             $\epsilon_f(\sim-\varepsilon_q)>\epsilon_F-D$ or
             $\epsilon_F+D>2\epsilon_f+U_f$, where charge
             fluctuations on the $d$ orbital is quenched but still
             exist on the $f$ orbital. The system is in the single
             impurity ($SU(2)$)  Kondo impurity regime;
\item[(\emph{iii})] $\epsilon_F-D>\epsilon_f$ and
             $\epsilon_F+D<2\epsilon_f+U_f$,  but $D>|J_{df}|$ where charge
             fluctuations in both $d$ and $f$ orbitals are quenched,  but
             the singlet-triplet energy splitting can be
             neglected. The singlet and triplet states can be
             considered as degenerate and the system demonstrate
             the SO(4) Kondo regime.
\item[(\emph{iv,v})] The system is at the $SO(3)$ Kondo regime [if
             $J_{df}<0$, interval ({\it{iv}})] or the
             self-screened Kondo regime [if $J_{df}>0$, interval
             ({\it{v}})]. These regimes exist only if
             $U_f>\epsilon_F-2\epsilon_f\sim$ a few
             $\varepsilon_q$.
\end{itemize}

\noindent The RG analysis for the various regimes now follows:

\noindent \underline{\text{\em Regime (i)}}: Charge fluctuations
on the $d$-orbital are integrated out as in
Ref.~[\onlinecite{Haldane-78}], but here the spin-singlet and
spin-triplet energies are,  generically, renormalized distinctly.
Renormalization of other quantities such as $V$, $\epsilon_f$ and
$V_{df}$ are weak and can be ignored. The scaling procedure of
$\varepsilon_{S(T)}$ then yields,%
\cite{Hewson-book,Kikoin-Avishai-02}
\begin{equation}
  \frac{d\varepsilon_{S(T)}}{d\ln{D}}=
  V_{S(T)}^2~
  \rho(\veps_F+D),
  \label{Haldane-scal}
\end{equation}
where $V_S^2=\alpha_S^2V_d^2$ and $V_T^2=V_d^2$. The difference
between $V_S^2$ and $V_T^2$ originates from the appearance of the
normalization factor $\alpha_S$ in the singlet state [see
(\ref{singlet-def})]. Notice that $V_{S}^2<V_{T}^2$ and the
triplet energy level renormalizes faster than the singlet level.
This opens the possibility that, as RG stops, the ground state of
the system may become a triplet if the single-triplet level
crossing occurs before quenching of charge fluctuations in the
$d$- and $f$-levels. Solving equation (\ref{Haldane-scal}), we
find that the singlet-triplet energy spacing
$J_{df}^{(i)}(D)=\varepsilon_{T}(D)-\varepsilon_{S}(D)$ is given
by
\begin{eqnarray}
  J_{df}^{(i)}(D)
  \sim
  \beta_f^2\Delta_f\left(1-
       \frac{2V_d^2\rho_0}{\Delta_f}~
       \frac{\sqrt{D_0}-\sqrt{D}}
            {\sqrt{\varepsilon_0}}\right).
  \label{tilde-J-def}
\end{eqnarray}
where $D=D_i\sim\epsilon_F-\epsilon_d$ at the end of scaling.

\begin{figure}[htb]
\centering
\includegraphics[width=80mm, angle=0]
  {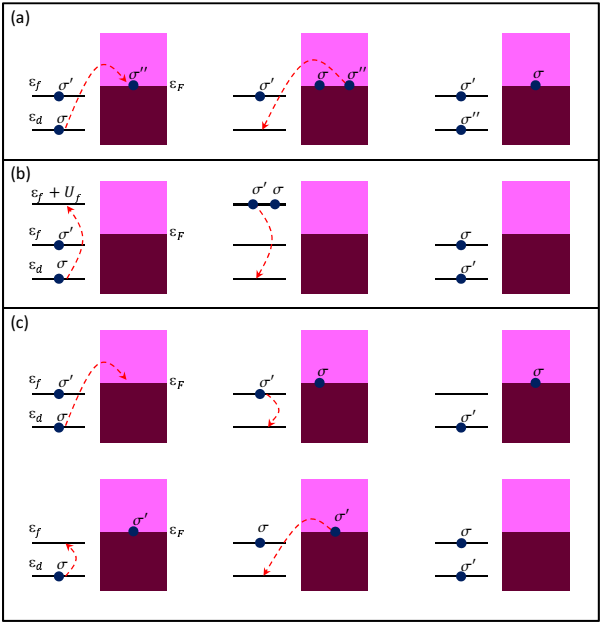}
 \caption{\footnotesize Transitions induced by $H_K^{(ii)}$,
 equation (\ref{sw1k}) leading from an initial state on the left
 and ending at a final state on the right. The various processes
 are associated with exchange constants (a): $J_{\gamma \gamma}$,
 (b): $J_{ff}$ and (c): $J_{\gamma f}$ and $J_{f \gamma}$}.
 \label{Fig-H-SU2}
\end{figure}

\noindent \underline{\text{\em Regime (ii)}}: Charge fluctuations
of the $d$-level are quenched at $D_i\sim\epsilon_F-\epsilon_d$. A
Schrieffer-Wolf transformation for the d-level, yields an
effective Hamiltonian
$$
  H^{(ii)}_I=H_0+H_f+H_{df}+H_K^{(ii)},
$$
where
\begin{equation}
  \label{sw1k}
  H_K^{(ii)} =
  {1\over2}
  \sum_{\alpha_\sigma, \beta_\sigma=\gamma_\sigma,f_\sigma}
  J_{\alpha\beta}
  \left(
       \alpha^{\dagger}_{\sigma}
       {{\mathbf{s}}_{\sigma\sigma'}}
       \beta_{\sigma'}
  \right)
  \cdot
  \mathbf{S}_d,
\end{equation}
where
$$
  \gamma_{\sigma}=
  \sum_{\nu,\mathbf{k}}
  \gamma_{\nu\mathbf{k}\sigma},
$$
${\mathbf{s}}$ is the vector of Pauli matrices, $\mathbf{S}_d$ is
the localized spin of the $d$-level and
\begin{eqnarray*}
  &&
  J_{\gamma \gamma}=
  \frac{2 V_d^2}{\epsilon_F-\epsilon_d},
  \ \ \ \ \
  J_{ff}=
  \frac{2 V_{df}^2}{\epsilon_f-\epsilon_d},
  \\
  &&
  J_{\gamma f}=
  J_{f \gamma}=
  \frac{V_dV_f} {\epsilon_f-\epsilon_d}+
  \frac{V_dV_f} {\epsilon_F-\epsilon_d}.
\end{eqnarray*}
The Hamiltonian   $H_K^{(ii)}$ describes coupling of the $d$-spin
to $\gamma$ and $f$ electrons. Notice that charge fluctuations in
the $f$-orbital is allowed in $H_K^{(ii)}$ through the mixed spin
operators
$$
  \mathbf{S}_{m}=
  \frac{1}{2}~
  \left(
       \gamma^{\dag}_{\sigma}
       {\mathbf{s}_{\sigma\sigma'}}
       f_{\sigma'}
  \right)
$$
and $\mathbf{S}_{m}^{\dag}$. Figure \ref{Fig-H-SU2} clarifies the
physics of the various interactions encoded in $H_{K}^{(ii)}$.

Using poor-man scaling,\cite{Anderson-70} charge fluctuations in
the $f$-orbital can now be integrated out. The dimensionless
exchange coupling constants $j_{\alpha \beta} \equiv \rho_0
J_{\alpha\beta}$ are renormalized as,
\begin{equation}
  \frac{dj_{\alpha\beta}}{d\ln{D}}=
       -j_{\alpha\gamma}j_{\gamma\beta}~
  \frac{\bar{\rho}}
       {\rho_0},
  \label{pmscaling}
\end{equation}
with the solution (see equation (\ref{rhoav}) for ${\cal
L}(D,D_i)$),
\begin{eqnarray}
  j_{\alpha\beta}(D) =
  j_{\alpha\beta}(D_i)+
  \frac{j_{\alpha\gamma}(D_i)
        j_{\gamma\beta}(D_i)
        {\cal{L}}(D,D_i)}
       {\displaystyle
        1-
        j_{\gamma\gamma}(D_i)
        {\cal{L}}(D,D_i)}.
  \label{J(T)-SO(3)}
\end{eqnarray}
The scaling invariant, that is, the Kondo temperature
$T_K^{(ii)}$, is got by solving the equation
\begin{equation}
  j_{\gamma\gamma}(D_i){\cal{L}}(T_K^{(ii)},D_i)=1,
  \label{TK-SU(2)}
\end{equation}
and scaling stops if
$$
  T_K^{(ii)} >
  D_{ii}=
  \min
  \big\{
      \epsilon_F-\epsilon_f,
      U_f-\epsilon_F+\epsilon_f
  \big\},
$$
at which the $d$-spin is quenched by the KE. This happens if
$j_{\gamma\gamma}(D_i)$ is large enough. We note that
${\cal{L}}(D_{ii},D_i)$ remains finite as long as
$D_{ii}>T_K^{(ii)}$. For $T_K^{(ii)}<D_{ii}$, scaling stops at
$D=D_{ii}$ and the $d$-spin is not quenched. The effective
coupling between $f$- and $d$-spins is a sum of contributions from
$H_{df}$ and $H_K^{(ii)}$, i.e.
$J_{df}=J_{df}^{(i)}(D_i)+J_{ff}(D_{ii})$. It is ferromagnetic if
$J_{df}<0$. Assuming that
$1-j_{\gamma\gamma}(D_i){\cal{L}}(D_{ii},D_i)$ is of order $O(1)$,
$\epsilon_F\sim\varepsilon_q<\varepsilon_0\ll|\epsilon_d|$ and
$U_f$ is a few $\veps_q$ [$\varepsilon_q\ll|\epsilon_d|$], we find
that $J_{ff}(D_{ii})$ is of the same order as $\beta_f^2\Delta_f$
and $J_{df}<0$ if
$$
  |V_d|^2 \geq
  \veps_0^{3\over2}~
  \sqrt{|\epsilon_d|}.
$$

\noindent \underline{\text{\em Regimes (iii) and (iv)}}: The
scaling stops at regime (ii) with
$D_{ii}\sim\epsilon_F+\varepsilon_q$ if the chemical potential is
located inside the band-gap which is the usual case for
insulators. A more interesting scenario occurs if the chemical
potential $\epsilon_F$ is located above the band-gap which may
happen if the insulator is doped by impurities. Scaling continues
in this case where charge fluctuations in $f$-level is also
quenched. In this case the mixed spin term $\mathbf{S}_m$ becomes
ineffective and we are left with an effective Hamiltonian
$H^{(2)}=H_0+H_K^{(iii)}$, where
\begin{equation}
  \label{sw2k}
  H_K^{(iii)}=
  J_K
  \mathbf{S}_{\gamma}
  \cdot
  \mathbf{S}_d+
  J_{df}
  \mathbf{S}_f
  \cdot
  \mathbf{S}_d,
\end{equation}
where
$$
  \mathbf{S}_{\gamma}=
  {1\over2}
  \sum_{\mathbf{k}\sigma,\mathbf{k}'\sigma',\nu}
  \left(
       \gamma^{\dag}_{\nu\mathbf{k}\sigma}
       {\mathbf{s}}_{\sigma\sigma'}
       \gamma_{\nu\mathbf{k}'\sigma'}
  \right),
$$
and
$$
  J_K=
  J_{\gamma\gamma}(D_{ii})=
  \frac{J_{\gamma\gamma}(D_i)}
       {1-
        j_{\gamma\gamma}(D_i)
        {\cal{L}}(D_{ii},D_i)}.
$$
Notice that $|J_{df}|$ is in general smaller than $J_K$.\cite{24}

\noindent \underline {\it $SO(3)$ and $SO(4)$ fixed points}:  For
$D_{ii}\gg|J_{df}|$, there exists a regime $D\gg|J_{df}|$ where
the system is governed by a critical point between the $SO(3)$-
and the quenched-Kondo regimes which has $SO(4)$ symmetry.%
\cite{27} In this regime the behavior of the system is governed by
the $SO(4)$ critical point. The system crossovers to the low
temperature $SO(3)$- or self-screened-Kondo regime at
$D<|J_{df}|$. We first consider the $SO(4)$ regime $D\gg|J_{df}|$.

\noindent \underline{\text{\em SO(4)  Kondo fixed point}}: In this
case we may set $J_{df}=0$ and apply the Schrieffer-Wolf
transformation directly to the spin-singlet and triplet states to
get,
\begin{eqnarray}
  H_K^{SO(4)} &=&
  J_{T}
  \big(
      {\bf{S}}
      \cdot
      {\bf{S}_{\gamma}}
  \big)+
  J_{ST}
  \big(
      {\bf{R}}
      \cdot
      {\bf{S}_{\gamma}}
  \big),
  \label{HK-singlet-triplet}
\end{eqnarray}
where ${\bf{S}}$ and ${\bf{R}}$ are the ($S=1)$ spin and the
Runge-Lenz operators, respectively that are expressible in terms
of Hubbard operators for the composed impurity, and satisfy the
$so(4)$ algebra.\cite{28} The exchange constants $J_{T}=J_K/2$ and
$J_{ST}=\alpha_S J_{T}$ scale as,
\begin{subequations}
\begin{eqnarray}
  \frac{dj_{T}}{d\ln{D}} &=&
  -\big(
       j_{T}^2+
       j_{ST}^2
   \big)~
  \frac{{\bar \rho}}
       {\rho_0},
  \label{scaling-T}
  \\
  \frac{dj_{ST}}{d\ln{D}} &=&
  -2
   j_{T}
   j_{ST}~
  \frac{{\bar \rho}}
       {\rho_0}.
  \label{scaling-ST}
\end{eqnarray}
  \label{subeqs-scaling-T-ST}
\end{subequations}
The combinations $j_{n}=j_T-(-1)^{n}j_{ST}$ [$n=1,2$] satisfy,
\begin{eqnarray}
  j_n(D) &=&
  \frac{j_n(D_{ii})}
       {\displaystyle
        1-
        j_n(D_{ii})~
        {\cal{L}}(D,D_{ii})},
  \label{Jn(T)-SO(4)}
\end{eqnarray}
whence
\begin{subequations}
\begin{eqnarray}
  &&
  j_{T}(D) ~=~
  \frac{1}{2}~
  \Big[
      j_1(D)+
      j_2(D)
  \Big],
  \label{J(T)-SO(4)}
  \\
  &&
  j_{ST}(D) =~
  \frac{1}{2}~
  \Big[
      j_1(D)-
      j_2(D)
  \Big].
  \label{J(ST)-SO(4)}
\end{eqnarray}
\end{subequations}
The corresponding Kondo temperature $T_{K_4}$ is determined from
the equation,
\begin{eqnarray}
  \Big\{
      j_{T}(\bar{D}_{ii})+
      j_{ST}(\bar{D}_{ii})
  \Big\}~
  {\cal{L}}(T_{K_4},\bar{D}_{ii})
  &=&
  1
  \label{TK-SO(4)}
\end{eqnarray}
provided $T_{K_4}\gg|J_{df}|$. For  $0<T_{K_4}<J_{df}$  the two
spins form a spin-singlet (self-screened KE) at $D\sim J_{df}$.

\noindent \underline{\text{\em SO(3) Kondo fixed point}}: For
$J_{df}<0$ and $T_{K_4}<|J_{df}|$, renormalization of $j_{ST}$
stops at $D=D_{iii}\sim|J_{df}|$. For $D<D_{iii}$, the Kondo
Hamiltonian becomes
\begin{eqnarray}
  H_{K}^{SO(3)} &=&
  J_{T}{\bf{S}}
  \cdot
  {\bf{S}_{\gamma}}.
  \label{HK-triplet}
\end{eqnarray}
The scaling equation for $j_T=J_T\rho_0$ and its solution are,
\begin{subequations}
\begin{eqnarray}
  &&
  \frac{dj_{T}}{d\ln{D}} =
  -j_{T}^2~
  \frac{{\bar \rho}}
       {\rho_0},
  \label{scaling-eq-T-SO3}
  \\
  &&
  j_{T}(D) =
  \frac{j_{T}(D_{iii})}
       {\displaystyle
        1-
        j_{T}(D_{iii})
        {\cal{L}}\big(D,D_{iii}\big)}.
  \label{scaling-sol-T-SO3}
\end{eqnarray}
  \label{subeqs-SO3-scaling}
\end{subequations}
Scaling stops at $T_{K_3}$ determined from the equation,
\begin{eqnarray}
  j_{T}\big(D_{iii}\big)~
  {\cal{L}}\big(T_{K_3},D_{iii}\big)
  &=&
  1.
  \label{TK-SO(3)}
\end{eqnarray}

\section{Resistivity and Impurity Magnetic Susceptibility}
  \label{sec-resist-suscept}

Having elaborated upon the theory in the weak coupling regime $T
\gg T_K$'s we are now in a position to carry out perturbation
calculations of experimental observables. In 3D, the most
accessible ones are the impurity resistivity $R_{\rm{imp}}(T)$ and
the impurity magnetic susceptibility $\chi_{\rm{imp}}(T)$. We
shall be guided by the quest to find out how the special features
of the TI's are reflected in these observables. These features are
the occurrence of gap and the structure of the DOS especially near
the band edges $\pm\varepsilon_q$. In addition, reducing the
temperature results in the crossover between different scaling
regimes of the couplings. Explicitly, there are three relevant
temperature regimes denoted as {\it (ii),(iii),(iv)} in order to
match the notation of the corresponding scaling regimes discussed
previously. The first regime, denoted as {\it (ii)}, is defined by
[$D_i>T>D_{ii}$] as given by equation (\ref{pmscaling}) for the
scaling interval (\emph{ii}). Local moment behavior exists only at
the $d$-level in this regime and therefore there is Kondo
scattering with SU(2) symmetry. The second regime, denoted as
({\it{iii}}), is defined by equation (\ref{subeqs-scaling-T-ST})
for scaling interval (\emph{iii}) [$D_{ii}>T>D_{iii}$]. Here one
may neglect the difference in energies between the singlet and
triplet states and the system is at the $SO(4)$ Kondo regime. The
third regime, denoted as {\it (iv)}, is defined by equation
(\ref{scaling-eq-T-SO3}) for scaling interval (\emph{iv})
[$D_{iii}>T>T_{K_3}$].  Here there is Kondo scattering with
$SO(3)$ symmetry (when $J_{df} <0$) or a self-screened KE if
$J_{df}>0$. The temperature dependence of the resistivity and
magnetic susceptibility in these three different scaling regimes
are distinct.

\noindent In the calculation of resistivity, we assume
$\epsilon_F>\varepsilon_q$ (the TI is doped) and the system has a
Fermi surface. The impurity resistivity as calculated in the
framework of the ``poor man's scaling'' formalism is given by,
\begin{eqnarray}
  R_{\rm{imp}} &=&
  \frac{N_{\nu}R_0}
       {{\cal{L}}^{2}(T_{K_{\nu}},T)},
  \label{resist}
\end{eqnarray}
where $\nu=(ii),(iii),(iv)$, denotes the pertinent temperature
regime as detailed above. The corresponding Kondo temperatures are
$T_{K_{ii}}$, equation (\ref{TK-SU(2)}),
$T_{K_{iii}}\equiv{T}_{K_4}$, equation (\ref{TK-SO(4)}), or
$T_{K_{iv}}\equiv{T}_{K_3}$, equation (\ref{TK-SO(3)}). The
numerical factors $N_{\nu}$ are, $N_{ii}=N_{iii}=3/4$ and
$N_{iv}=2$. Here
\begin{eqnarray*}
  R_0 =
  \frac{3 \pi c_{\rm{imp}}}
       {\hbar e^2 \rho_0^2}~
  \frac{1}
       {v_1^2+v_2^2},
  \ \ \ \ \
  v_i \approx
  \frac{1}{\hbar}~
  \bigg(
       \frac{\partial\veps_{k_i}}
            {\partial k_i}
  \bigg),
\end{eqnarray*}
and $k_1,~k_2$ are two solutions of $\veps_k=\epsilon_F$ (see
Figure \ref{Fig1}).

\begin{figure}[htb]
\centering
\includegraphics[width=60mm,angle=0]
  {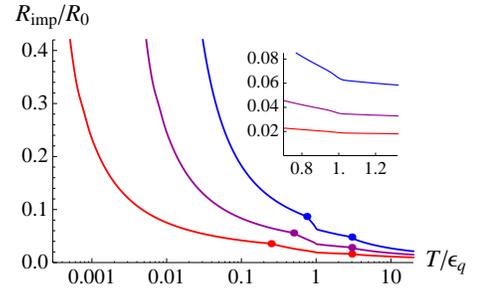}
 \caption{\footnotesize
   Resistivity (\ref{resist}) as a function of temperature for
   $\epsilon_d=-80\veps_q$, $\veps_0=24\varepsilon_q$,
   $\epsilon_F=2\veps_q$ and different values of $j$:
   $j=0.1$ [bottom red curve], $j=0.12$ [middle purple curve]
   and $j=0.14$ [top blue curve]. The dots denote $T=D_{ii}$
   and $T=D_{iii}$ separating the temperature intervals
   (\emph{ii}), (\emph{iii}) and (\emph{iv}). Inset: behavior
   of $R_{\rm{imp}}$ for the temperature $T\sim\veps_q$.}
 \label{Fig-R(T)}
\end{figure}

The resistivity as a function of the temperature is shown in
Fig.\ref{Fig-R(T)} assuming a low temperature $SO(3)$ fix point.
It is seen that $R_{\rm{imp}}$ has different temperature
dependence within the temperature intervals (\emph{ii}),
(\emph{iii}) and (\emph{iv}), with crossover observed at
$T=D_{ii}$ and $T=D_{iii}$ [the points $D_{ii}$ and $D_{iii}$ are
denoted by dots].  In addition, crossovers are observed at
$T=\veps_q$ [interval (\textsl{iii})]. These crossovers appear
since the function ${\cal{L}}(T_{K_{iii}},T)$ changes its behavior
at $T=\epsilon_F-\veps_q=\veps_q$ [we take $\epsilon_F=2\veps_q$
here].

The Kondo scattering manifests itself also in the magnetic
susceptibility.\cite{Hewson-book} The impurity susceptibility
calculated in the framework of the ``poor man's scaling'' is
\begin{eqnarray}
  \chi_{\rm{imp}} &=&
  \frac{K_{\nu}\chi_0T_{K_{ii}}}{T}~
  \bigg\{
       P_{\nu}-
       \frac{1}{{\cal{L}}(T_{K_{\nu}},T)}
  \bigg\},
  \label{suscept}
\end{eqnarray}
where $\nu=(ii),(iii),(iv)$, the Kondo temperatures are
$T_{K_{ii}}$, equation (\ref{TK-SU(2)}),
$T_{K_{iii}}\equiv{T}_{K_4}$, equation (\ref{TK-SO(4)}), and
$T_{K_{iv}}\equiv{T}_{K_3}$, equation (\ref{TK-SO(3)}). The
numerical factors are $K_{ii}=K_{iii}=1/4$, $K_{iv}=2/3$,
$P_{ii}=P_{iv}=1$ and $P_{iii}=2$. The constant $\chi_0$ is
$$
  \chi_0 =
  \frac{4c_{\rm{imp}} \mu_B^2}{T_{K_{ii}}}.
$$

\begin{figure}[htb]
\centering
\includegraphics[width=60mm,angle=0]
  {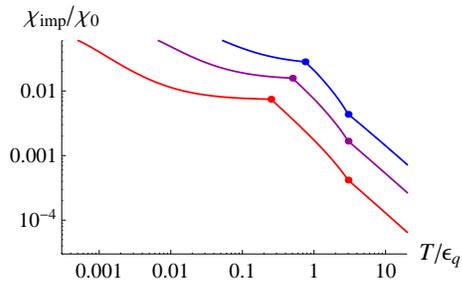}
 \caption{\footnotesize
   Magnetic susceptibility (\ref{suscept}) as a function of
   temperature for $\epsilon_d=-80\veps_q$,
   $\veps_0=24\varepsilon_q$, $\epsilon_F=2\veps_q$ and different
   values of $j$: $j=0.1$ [red curve], $j=0.12$ [purple curve] and
   $j=0.14$ [blue curve]. The dots denote $T=D_{ii}$ and
   $T=D_{iii}$ separating the temperature intervals (\emph{ii}),
   (\emph{iii}) and (\emph{iv}).}
 \label{Fig-chi(T)}
\end{figure}

The impurity magnetic susceptibility as a function of $T$ is shown
in Figure \ref{Fig-chi(T)}. The different temperature dependences
of $\chi_{\rm{imp}}$ at different temperature regimes (\emph{ii}),
(\emph{iii}) and (\emph{iv}) are obvious, with crossovers observed
at $T=D_{ii}$ and $T=D_{iii}$ [the points $D_{ii}$ and $D_{iii}$
are denoted by dots].

\section{Conclusions}
  \label{sec-conclusions}

We have analyzed the interplay between the Anderson impurity and
its induced in-gap bound state in a model of 2D topological
insulator. Using a weak-coupling RG analysis, it is shown that the
exchange interaction $J_{df}$ between the $d$- and the induced
in-gap $f$-spins may be renormalized dynamically to either
positive or negative values. The parameters required to observe
the above phenomena is not too restrictive
($|V_d|^2\geq\veps_0^{3\over2}\sqrt{|\epsilon_d|}$,
$\veps_q<U_f\ll{U}_d$) and is realistic.  The system exhibits
complex crossover behaviors at different parameter regimes as a
result which can be observed in the temperature dependence of the
impurity induced resistance and magnetic susceptibility. The
crossover in the temperature dependence of both the resistivity
and the impurity magnetic susceptibility at different regimes is a
peculiar feature that can serve as an experimental confirmation of
the above analysis. For both screened and under-screened Kondo
effect in the weak coupling regime, the effective coupling
constant $j$ renormalizes as $1/{\cal{L}}(T_K,T)$ (or, as
$1/\ln(T/T_K)$, when the DOS is flat, see Ref.
[\onlinecite{Hewson-book}]). As a result, the impurity
resistivity, $R_{\rm{imp}}$, behaves as $1/{\cal{L}}^2(T_K,T)$
[see equation (\ref{resist})], whereas the susceptibility,
$\chi_{\rm{imp}}$, is given by equation (\ref{suscept}).

\noindent The physics described above is not limited to TI but is
a general consequence of (doped) insulators (and semi-conductors)
with a large electronic density of states at the band edge such
that in-gap bound states are easily induced by an Anderson
impurity. Similar physics may be found in for example, two-layer
graphene systems. Our paper is just a first step towards
understanding the rich physics associated with impurities in these
systems.

\noindent{\bf Acknowledgements}: Discussions with C. M. Varma are
highly appreciated. We acknowledge support by HKRGC through grant
HKUST03/CRF09. The research of I.K and Y.A is partially supported
by grant 400/12 of the Israeli Science Foundation (ISF).

\appendix

\section{Interpretation of $V_0$ in equation (\ref{V-def})}
  \label{append-what-is-V0}

\begin{figure}[htb]
\centering
\includegraphics[width=60 mm, angle=0]
  {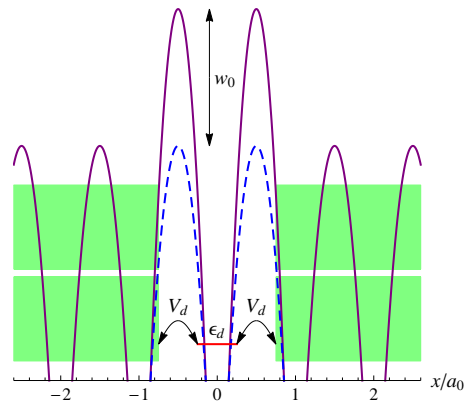}
 \caption{\footnotesize
 Potential energy of electrons in the lattice with the impurity
 put at the point ${\bf{r}}={\bf{0}}$. The solid purple curve is
 the potential energy (\ref{V-lattice-imp}), whereas the dashed
 blue curve is the potential energy (\ref{V-lattice}) of electrons
 in the lattice without impurity. The filled areas denote the
 valence and conduction bands. The red line is the impurity level.
 It is assumed that $|{\bf{a}}_{1,2,3}|=a_0$ and ${V_d}\ll{w_0}$,
 where $V_d$ is the hybridization rate between $d$-impurity and
 the band electrons.}
 \label{Fig-what-is-V}
\end{figure}

Electrons in an lattice move in the periodic potential
$w_{l}({\bf{r}})$,
\begin{eqnarray}
  w_l({\bf{r}}) &=&
  \sum_{\bf{n}}
  w_a({\bf{r}}-{\bf{n}}),
  \label{V-lattice}
\end{eqnarray}
where $w_a({\bf{r}})$ is the interaction energy of electrons with
an lattice atom,
${\bf{n}}=n_1{\bf{a}}_1+n_2{\bf{a}}_2+n_3{\bf{a}}_3$,
${\bf{a}}_{1,2,3}$ are the lattice vectors, $n_{1,2,3}$ are
integers. In this case, electrons tunnel from one atom to another
and the single-electron atomic levels $\epsilon_a$ reduce to the
energy band shown in Figure \ref{Fig-what-is-V}.

When the atom of the lattice at the point ${\bf{r}}={\bf{0}}$ is
replaced by an impurity atom, the potential energy $w_i({\bf{r}})$
of interaction of electrons with the impurity differs from
$w_a({\bf{r}})$. As a result, the potential energy of electrons in
the lattice with the impurity,
\begin{eqnarray}
  w({\bf{r}}) &=&
  \sum_{{\bf{n}}\ne0}
  w_a({\bf{r}}-{\bf{n}})+
  w_i({\bf{r}}),
  \label{V-lattice-imp}
\end{eqnarray}
is not periodic anymore (see the purple curve in Figure
\ref{Fig-what-is-V}).

Then the potential scattering can be estimated as,
\begin{eqnarray}
  V_0 &=&
  \int{d^3{\bf{r}}}~
  \Big\{
      w_i({\bf{r}})-
      w_a({\bf{r}})
  \Big\} \sim
  \nonumber \\ &\sim&
  w_0~
  a_0^3,
  \ \ \ \ \
  w_0=w_i^{(0)}-w_a^{(0)},
  \label{V0-estimate}
\end{eqnarray}
where $w_i^{(0)}$ or $w_a^{(0)}$ is the peak of the potential
energy of the impurity or the atom. Here we assume that the
electric potential is screened at the inter-atomic distance $a_0$.

\section{Local Density of States}
  \label{append-local-DOS}

The potential scattering $V$, equation (\ref{V-def}), results in
modification of the density of states and formation of an in-gap
energy level . In order to derive an explicit expression for the
local DOS (\ref{DOS-local-def}), we calculate the retarded Green's
function (\ref{G-nuk-veps-def}). Applying the equation of motion
(\ref{eq-Heisenberg}), we get
\begin{widetext}
\begin{eqnarray}
  G_{\nu{\bf{k}},\nu'{\bf{k}}'}(\veps)
  &=&
  g_{\nu k}(\veps)~
  \delta_{kk'}~
  \delta_{\nu\nu'}+
  V_0~
  g_{\nu k}(\veps)
  \sum_{\nu''{\bf{k}}''}
  G_{\nu''{\bf{k}}'',\nu'{\bf{k}}'}(\veps),
  \label{eq-motion-GF}
\end{eqnarray}
\end{widetext}
where
\begin{eqnarray*}
  &&
  g_{\nu k}(\veps) =
  \frac{1}{\veps-\nu\veps_{k}+i\eta}.
\end{eqnarray*}
The solution of equation (\ref{eq-motion-GF}) is,
\begin{eqnarray}
  G_{\nu{\bf{k}},\nu'{\bf{k}}'}(\veps) &=&
  g_{\nu k}(\veps)~
  \delta_{kk'}~
  \delta_{\nu\nu'}+
  \nonumber \\ &+&
  \frac{V_0~
        g_{\nu k}(\veps)~
        g_{\nu'k'}(\veps)}
       {\displaystyle
        1-
        V_0
        \sum_{\nu''{\bf{k}}''}
        g_{\nu''k''}(\veps)}.
  \label{GF-kk'-res}
\end{eqnarray}
Then the DOS (\ref{DOS-local-def}) is
\begin{eqnarray}
  \rho(\veps,r) =
  \rho_{0}(\veps)-
  {\rm{Im}}
  \frac{\pi V_0
        \big\{
            {\cal{R}}^2(\veps,r)+
            \pi^2
            \rho_0^2(\veps,r)
        \big\}}
       {1+
        \pi V_0
        \big\{
            {\cal{R}}(\veps,0)+
            i\rho_0(\veps,0)
        \big\}},
  \label{DOS-GF-append}
\end{eqnarray}
where ${\cal{R}}(\veps,r)$ and $\rho_0(\veps,r)$ are real and
imaginary parts of the Green's function,
$$
  \tilde{g}(\veps,r) =
  -\frac{1}{\pi}
  \sum_{\nu{\bf{k}}}
  g_{\nu k}(\veps)
  e^{i{\bf{k}}{\bf{r}}}.
$$
Explicitly, $\rho_0(\veps,r)$ is given by equation
(\ref{DOS-bulk-res}) and
\begin{eqnarray*}
  {\cal{R}}(\veps,r) &=&
  -\frac{\Theta(\veps_q-|\veps|)}{\pi}
  \sum_{\nu{\bf{k}}}
  \frac{\sin\big(kr\big)}
       {k r~\big(\veps-\nu\veps_k\big)}.
\end{eqnarray*}

When $|\veps|>\veps_q$, ${\cal{R}}(\veps,r)$ vanishes and equation
(\ref{DOS-GF-append}) reduces to equation (\ref{DOS-local-res}).
When $|\veps|<\veps_q$, the bare DOS vanishes, but the DOS
(\ref{DOS-GF-append}) gets a delta peak due to the localized
$f$-level,
\begin{eqnarray}
  \rho(\veps,r) =
  \pi^2 V_0~
  {\cal{R}}^2(\epsilon_f,r)~
  \delta
  \big(
      1+
      \pi
      V_0
      {\cal{R}}(\veps,0)
  \big),
  \label{DOS-local-append}
\end{eqnarray}
where the condition of vanishing of the argument of the
delta-function gives us the secular equation (\ref{eq-secular})
for $\epsilon_f$. The amplitude ${\cal{R}}^2(\epsilon_f,r)$ of the
delta-peak vanishes when $r\to\infty$, so that $\epsilon_f$ is a
localized state.

\section{The $SO(4)$ Kondo Hamiltonian, equation
         (\ref{HK-singlet-triplet})}
  \label{append-HK-SO4}

The $SO(4)$ Hamiltonian is derived in regime (iii) of the RG
analysis when $\epsilon_F$ is located above the band-gap. It has
the form
\begin{eqnarray}
  H_K^{SO(4)} &=&
  J_{T}
  \big(
      {\bf{S}}
      \cdot
      {\bf{S}_{\gamma}}
  \big)+
  J_{ST}
  \big(
      {\bf{R}}
      \cdot
      {\bf{S}_{\gamma}}
  \big)
  \label{HK-singlet-triplet-append}
\end{eqnarray}
where ${\bf{S}}$ and ${\bf{R}}$ are the ($S=1)$ spin operator and
the Runge-Lenz operator, respectively with
\begin{eqnarray}
  &&
  S^{+}=
  \sqrt{2}
  \big(
      X^{10}+X^{0\bar{1}}
  \big),
  \ \ \ \ \
  S^{-}=
  \sqrt{2}
  \big(
      X^{01}+X^{\bar{1}0}
  \big),
  \nonumber \\ &&
  S^{z}=
  X^{11}-X^{\bar{1}\bar{1}},
  \label{S-vectors}
  \\
  &&
  R^{+}=
  \sqrt{2}
  \big(
      X^{1S}-X^{S\bar{1}}
  \big),
  \ \ \ \ \
  R^{-}=
  \sqrt{2}
  \big(
      X^{S1}-X^{\bar{1}S}
  \big),
  \nonumber \\ &&
  R^{z}=
  -\big(
      X^{0S}+X^{S0}
  \big).
  \label{R-vectors}
\end{eqnarray}
Here $X^{\lambda\lambda'}=|\lambda\rangle\langle\lambda'|$,
$|\lambda(\lambda')\rangle=|S\rangle,|T_m\rangle~(m=0,\pm1)$ are
the spin singlet and triplet states. The operators ${\bf S}$ and
${\bf R}$ are the generators of the group $SO(4)$, as they
satisfied the following commutation relations, ($i,j,k=x,y,z$,
summation convention implied),
\begin{eqnarray}
  \big[S_i,S_j\big] &=&
  i \veps_{ijk}S_k,
  \nonumber
  \\
  \big[R_i,R_j\big] &=&
  i \veps_{ijk}S_k,
  \label{SO4}
  \\
  \big[R_i,S_j\big] &=&
  i \veps_{ijk}R_k.
  \nonumber
\end{eqnarray}

\section{Resistivity}
  \label{append-resist}

The resistivity for the SU(2) symmetry [regime (ii)] calculated
within the third order of the perturbation theory
is\cite{Hewson-book}
\begin{eqnarray}
  R_{\rm{imp}}^{SU(2)} &=&
  \frac{3R_0}{4}~j_D^2+
  \frac{3R_0}{2}~j_D^3
  {\cal{L}}(T,\bar{D}),
  \label{resist-SU(2)-2nd-3rd}
\end{eqnarray}
where
\begin{eqnarray*}
  R_0 =
  \frac{3 \pi c_{\rm{imp}}}
       {\hbar e^2 \rho_0^2}~
  \frac{1}
       {v_1^2+v_2^2},
  \ \ \ \ \
  v_i \approx
  \frac{1}{\hbar}~
  \bigg(
       \frac{\partial\veps_{k_i}}
            {\partial k_i}
  \bigg),
\end{eqnarray*}
$k_1$ and $k_2$ are two solutions of the equation
$\veps_k=\epsilon_F$ (see Fig.1 of the main text). The function
${\cal{L}}(T_1,T_2)$ is given by equation (7) of the main text.

Applying the condition of invariance of the resistivity under the
``poor man's scaling'', we get
\begin{eqnarray}
  R_{\rm{imp}}^{SU(2)}(T) &=&
  \frac{3R_0}{4{\cal{L}}^2(T_{K_2},T)}.
  \label{R-SU(2)-poor-man-scal-res}
\end{eqnarray}
Here the factor $N_{\nu}=3/4$ comes from the factor $S(S+1)$ which
is $3/4$ for $S=1/2$.

The resistivity for the SO(4) symmetry [regime (iii)] calculated
within the third order of the perturbation theory is
\begin{eqnarray}
  R_{\rm{imp}}^{SO(4)} &=&
  \frac{3R_0}{2}~
  \Big(
      j_{T}^2+j_{ST}^2
  \Big)+
  \nonumber \\ &+&
  3R_0
  \Big[
      j_T
      \big(j_{T}^2+j_{ST}^2\big)+
      2j_T
      j_{ST}^2
  \Big]
  {\cal{L}}(T,\bar{D}).
  \nonumber \\
  \label{resist-SO(4)-2nd-3rd}
\end{eqnarray}
The couplings $j_{T}$ and $j_{ST}$ renormalize in such a way that
the difference $j_{T}-j_{ST}$ is finite (and small) even when the
temperature $T$ approaches the Kondo temperature ${T}_{K_4}$,
whereas $j_{T}+j_{ST}\to\infty$ when $T\to{T}_{K_4}$ [see
equations (16)--(18) in the main text]. As a result, the
resistivity for the SO(4) symmetry in the low-temperature regime
[$\bar{D}\gg{T}\gg{T}_{K_4}$] is described by equation
(\ref{R-SU(2)-poor-man-scal-res}).

For the SO(3) symmetry [regime (iv)], the resistivity calculated
within the third order of the perturbation theory is
\begin{eqnarray}
  R_{\rm{imp}}^{SO(3)} &=&
  2R_0j_T^2+
  4R_0j_T^3
  {\cal{L}}(T,\bar{D}).
  \label{resist-SO(3)-2nd-3rd}
\end{eqnarray}
Applying the condition of invariance of the resistivity under the
``poor man's scaling'', we get
\begin{eqnarray}
  R_{\rm{imp}}^{SO(3)}(T) &=&
  \frac{2R_0}{{\cal{L}}^2(T_{K_3},T)}.
  \label{R-SO(3)-poor-man-scal-res}
\end{eqnarray}
The factor $N_{\nu}=2$ comes from the factor $S(S+1)$ which is $2$
for $S=1$.

\section{Magnetic Susceptibility}
  \label{append-suscept}

The susceptibility for the SU(2) symmetry calculated within the
second order of the perturbation theory is \cite{Hewson-book}
\begin{eqnarray*}
  \chi_{\rm{imp}}^{SU_2}(T) &=&
  \frac{\chi_0T_{K_2}}{4T}
  \Big\{
       1-
       2J_D\rho_0-
       4J_D^2\rho_0^2
       {\cal{L}}(T,\bar{D})
  \Big\}.
\end{eqnarray*}
Applying the RG transformations, we get
\begin{eqnarray}
  \chi_{\rm{imp}}^{SU_2}(T) &=&
  \frac{\chi_0T_{K_2}}{4T}
  \bigg\{
       1-
       \frac{1}{{\cal{L}}(T_{K_2},T)}
  \bigg\},
  \label{suscept-SU(2)}
\end{eqnarray}
where the factor $K_{ii}=1/4$ comes from $S(S+1)/3$ which is $1/4$
for $S=1/2$.

For SO(4) symmetry, the impurity susceptibility calculated to the
second order in $J$'s is
\begin{eqnarray*}
  \chi_{\rm{imp}}^{SO_4}(T) &=&
  \frac{\chi_0T_{K_4}}{2T}
  \Big\{
       1-
       J_T\rho_0-
  \nonumber \\ && -
       \big(
           J_T^2\rho_0^2+
           J_{ST}^2\rho_0^2
       \big)
       {\cal{L}}(T,\bar{D})
  \Big\}.
\end{eqnarray*}
Applying the RG transformations, we get
\begin{eqnarray}
  \chi_{\rm{imp}}^{SO_4}(T) &=&
  \frac{\chi_0T_{K_4}}{4T}
  \Big\{
      2-
      \frac{1}{{\cal{L}}(T_{K_4},T)}
  \Big\}.
  \label{suscept-SO(4)}
\end{eqnarray}
The factors $P_{iii}=2$ and $K_{iii}=1/4$ have following origin:
there are two spins $S=1/2$ [so that $P_{iii}=2$], every spin
gives the factor $K_{iii}=S(S+1)/3=1/4$.

For SO(3) symmetry, the impurity susceptibility calculated to the
second order in $J$'s is
\begin{eqnarray*}
  \chi_{\rm{imp}}^{SO_3}(T) &=&
  \frac{2\chi_0T_{K_3}}{3T}
  \Big\{
       1-
       J_T\rho_0-
       J_T^2\rho_0^2
       {\cal{L}}(T,\bar{D})
  \Big\}.
\end{eqnarray*}
Applying the RG transformations, we get
\begin{eqnarray}
  \chi_{\rm{imp}}^{SO_3}(T) &=&
  \frac{2\chi_0T_{K_3}}{3T}
  \Big\{
      1-
      \frac{1}{{\cal{L}}(T_{K_3},T)}
  \Big\}.
  \label{suscept-SO(3)}
\end{eqnarray}
The factor $K_{iv}=2/3$ comes from $S(S+1)/3$ for $S=1$.


\end{document}